\documentclass[fleqn,10pt]{wlscirep}

\usepackage{graphicx}
\usepackage{color}
\usepackage{booktabs}

\newcommand\q{\phantom0}
\newcommand\qq{\q\q}
\newcommand\qc{\phantom{0,}}
\newcommand\qqc{\phantom{00,}}

\title{GDC 2: Compression of large collections of genomes} 
\author[1,*]{Sebastian Deorowicz}
\author[1]{Agnieszka Danek}
\author[2]{Marcin Niemiec}
\affil[1]{Institute of Informatics, Silesian University of Technology, Akademicka 16, 44-100 Gliwice, Poland}
\affil[2]{Nubitech, 40-684 Katowice, Poland}

\affil[*]{sebastian.deorowicz@polsl.pl}

\begin{abstract}
The fall of prices of the high-throughput genome sequencing changes the landscape of modern genomics.
A number of large scale projects aimed at sequencing many human genomes are in progress.
Genome sequencing also becomes an important aid in the personalized medicine.
One of the significant side effects of this change is a necessity of storage and transfer of huge amounts of genomic data.
In this paper we deal with the problem of compression of large collections of complete genomic sequences.
We propose an algorithm that is able to compress the collection of 1092 human diploid genomes about 9,500 times.
This result is about 4 times better than what is offered by the other existing compressors. 
Moreover, our algorithm is very fast as it processes the data with speed 200\,MB/s on a modern workstation.
In a consequence the proposed algorithm allows storing the complete genomic collections at low cost, e.g., the examined collection of 1092 human genomes needs only about 700\,MB when compressed, what can be compared to about 6.7\,TB of uncompressed FASTA files.
The source code is available at \url{http://sun.aei.polsl.pl/REFRESH/index.php?page=projects&project=gdc&subpage=about}.
\end{abstract}

\begin{document}
\flushbottom
\maketitle
\thispagestyle{empty}

\section*{Introduction}
The genome sequencing technology has recently become so cheap that it started to be considered as a useful tool in medicine.
Companies like Illumina offer whole human genome sequencing for medical purposes for five thousand U.S.\ dollars~\cite{Ill2015}.
There are also large scale projects designed to find the common differences between individual genomes.
One of the most famous is the 1000 Genome Project~\cite{TGP2012} which aims at sequencing the genomes of several thousand humans and determining the genetic variants with at least 1\% frequency.
There are, however, even broader attempts for human genome sequencing, to mention the UK10K project~\cite{UK10K}, the Personal Genomes Project~\cite{B2012}, and the Million Veteran Project (MVP)~\cite{MVP2013}.
The planned number of sequenced genomes are 10K, 100K, and 1M, respectively.
%The goal of the MVP is to sequence one million genomes.
Large collections of genomes are built also for other species.
E.g., in the 1001 Genomes Project (1001GP)~\cite{WM2009,1001GP} about 1000 of genomes of \emph{Arabidopsis thaliana} are to be sequenced.

The sequencing is of course challenging, but due to the large amounts of produced data, the pure storage and transfer of the results becomes a challenge too.
The recent papers~\cite{K2011,DG2013} show that the IT costs are (or will be soon) comparable to the sequencing costs.
Due to the slow progress in reducing the IT prices, the effective ways of representing genomic data in compact form are intensively investigated.
Several subproblems can be identified here. %, depending on the data type.
The first is the compression of raw sequencing reads~\cite{RD2014,BM2013,JRPK2012}
The second is the compression of reads after mapping onto reference genomes~\cite{FLC2011,JRPK2012,HNS2014}.
The third is the compression of results of variant calling~\cite{CLLX2009,PWY2013,DDG2013}.
The fourth is the compression of complete genomic sequences~\cite{DG2011,WL2012,WL2013,OHW2014}.
These subproblems are related, nevertheless require different approaches. 
The recent surveys discuss most of the existing algorithms~\cite{DG2013,GRU2014,ZZJHY2015}.
%, which are described in the mentioned surveys.

In this paper we deal with the last of the mentioned tasks, i.e., storage of collections of genomes. We propose Genome Differential Compressor 2 (GDC 2), a utility for compression of large sets of genomes of the same species. 
Since such genomes are highly similar, e.g., it was estimated that two humans have their genomes identical in 99.5 percent~\cite{L2007}, it is clear that when compressing a collection of genomes one can obtain better compression ratios than when compressing the sequences separately.
%There a number of works in this field.
Initially, the researchers tried to use the similarity between a sequence to be compressed and a reference sequence.
The first impressive result was by Christley \emph{et al.}~\cite{CLLX2009}.
They showed that the description of differences between James Watson's genome and the reference genome can be stored in as little as 4.1\,MB.
Taking into account that the complete haploid human genome is of size 3.1\,Gbases, this translates to $\sim$750-fold compression.
This result was recently improved by Pavlichin {\em et al.}~\cite{PWY2013} who reduced the space for the JW genome to about 2.5\,MB (compression ratio $\sim$1250).

Such large compression ratio was possible since the data were preprocessed, i.e., precise information of all variants were available.
This is not always the case, as the genomes can be obtained in different experiments with different reference genomes or the genomes can be \emph{de novo} assembled.
In such situations the data to be compressed are collections of complete genomic sequences.
This significantly complicates the compression task, as the differences between sequences are not given explicitly; they have to be found, e.g., by multiple complete genome alignment, which is a very complex problem.
Moreover, for technological reasons, the differences between \emph{de novo} assembled genomes are usually larger than between the reassembled genomes.

Several papers for the problem of compression of collections of genomic sequences were published~\cite{KPZ2011,OHW2014,PPG2012,WL2012,DG2011}.
In majority of them, each single sequence is compressed separately, by identifying the differences between it and a single reference genome.
This allowed to obtain compression ratios for human genomes up to 400, much poorer than $\sim$1250 obtained by Pavlichin {\em et al.}~\cite{PWY2013}.
This is the price for the lack of prior knowledge about the compressed data.
The most successful attempts at obtaining higher compression ratios were possible by exploring the knowledge of similarities between more sequences in the collection.
Since such approaches are the real competitors to the proposed algorithm, we will describe them a little more.

The first attempt in this direction was GDC-ultra~\cite{DG2011}.
It takes a single reference sequence and constructs a search structure (namely, hash table) for it.
Then it compresses the first sequence of the collection by looking for similarities between this sequence and the reference.
When the sequence is processed, it is used as an additional reference sequence for further sequences, so a separate search structure is constructed for it.
The same is for the following sequences, so for example, the 25th input sequence of the collection is compressed by looking for the differences between it and: the main reference sequence, the formerly processed 24 sequences of the collection.
The number of additional reference sequences is limited to 39 (for technical reasons only, mainly to keep the necessary amount of memory at a reasonable level).
If the collection consists of more than 39 sequences, the 40th, 41st, etc. sequence is compressed with the 40 references only.
The differences between the current sequence and the referential sequences is finally Huffman coded.
Such approach proved to be promising, since the collection of 69 human genomes were compressed with ratio $\sim$1000.

A different approach was used by Wandelt \emph{et al.}~\cite{WL2013} in their FRESCO algorithm.
They investigated several variants, and below we will describe the one that gave the best results.
The collection is divided into two sets: (\emph{i}) additional references, (\emph{ii})~remaining sequences.
FRESCO constructs a search structure (suffix tree) for the main reference sequence.
Then it looks for similarities between the additional reference sequences and the main reference performing classical Ziv--Lempel parsing of additional reference sequences.
As a results it obtains for each additional reference a sequence of triples (position in the main reference, length of the identical part, next symbol).
For the Ziv--Lempel-parsed additional reference sequences a search structure (hash table) is built.
After that FRESCO is ready to perform the compression of the remaining sequences from the collection.
Each sequence is Ziv--Lempel-parsed against the main reference sequence.
Then, the sequence of triples is compressed using the additional Ziv--Lempel-parsed reference sequences serving as the second level reference.
The obtained compression ratios are impressive as they are approximately 3000 for the collection of about 1000 haploid genomes of the 1000GP, when 70 additional reference sequences were used.
 
The best compression ratios for the genomic collection was obtained by TGC algorithm~\cite{DDG2013}.
It is, however, from a different category, since as an input it takes a Variant Call Format (VCF)~\cite{D2011} file describing the differences between genomes and the reference sequence, so it processes essentially the same data as Pavlichin {\em et al.}~\cite{PWY2013}.
In this work we deal with complete genomes stored in FASTA format.
In theory it is possible to convert FASTA files into VCF files, but it would require making a perfect alignment of many complete genomes, which is far from being trivial, especially due to a presence of long structural variants.
Nevertheless, comparing the obtained results with TGC will be interesting, as it will allow us to see how far we are from the top algorithm for the similar problem.
The main idea of TGC is to split the VCF file into two files.
The first (dictionary of variants) stores a description of each variant (i.e., its type, position, alternative alleles, etc.).
The second file stores the binary representation of presence/absence of each single variant in each single sequence.
The bit vectors (one for each individual) are compressed using a specialized Ziv--Lempel-based algorithm.
The dictionary file is also compressed using a specialized algorithm.
The compression ratios of TGC for the collection of 1092 diploid human genomes (when taking only 1 reference sequence) is about 15,500.

\section*{Methods}
\subsection*{Definitions}
For precise description of the proposed algorithm let us define some terms.
As an input we have a single reference sequence~$R$ and a collection of genome sequences $\mathcal{S} = \{S^1, S^2, \ldots, S^n\}$.
Each sequence is composed of symbols from some alphabet $\Sigma$, i.e, $S^k = s_1^k s_2^k \ldots s_{|S^k|}^k$ for each $1 \le k \le n$, where $s_i^k \in \Sigma$ for each valid~$i$ and $|S^k|$ denotes the length of $S^k$.
Also $R = r_1 r_2\ldots r_{|R|}$, where $r_i \in \Sigma$  for each valid~$i$ and $|R|$ denotes the length of $R$.
For any sequence $X$ (a reference or from the collection) $X_{i,j} = x_i x_{i+1} \ldots x_j$.

For the DNA sequences the alphabet should ideally contain only 4~symbols (\textsf{A}, \textsf{C}, \textsf{G}, \textsf{T}), but in practice \textsf{N} (unknown) symbols are quite frequent.
Moreover, sometimes also other IUPAC codes appear.
Thus in the work we assume only that the symbols are letters from the ASCII code (we also distinguish between lower- and uppercase letters).

\subsection*{Compression algorithm}
At the beginning, the compression algorithm reads the reference sequence~$R$ and constructs a search structure $\mathit{HT}^\text{R}$ (namely, hash table with linear probing) for it.
The hash value is computed for each $h_{1m}$-symbol long substring of $R$ ($h_{1m} = 15$ by default, but a different value can be specified by a user), i.e., for all $R_{i, i+h_{1m}-1}$, where $1 \le i \le |R|-h_{1m}+1$.
After that, the main processing of the collection $\mathcal{S}$ starts.
The compression algorithm is two-level.

At the first level, we perform the Ziv-Lempel factoring of all sequences from the collection~$\mathcal{S}$.
This means that for each sequence $S^k$ from $\mathcal{S}$ we produce a sequence~$L^k$ composed of tuples.
To this end, we start from $i=1$ and look for the longest common substring $S^k_{i,j}$ present in $R$.
Since the search structure $\mathit{HT}^\text{R}$ contains substrings of length $h_{1m}$ it is not possible to find shorter matches.
There are two possibilities here:
\begin{itemize}
\item No match of length at least $h_{1m}$ is found. 
Then, we append a tuple describing single symbol $s^k_i$, i.e., $\langle f_\text{literal}, s^k_i\rangle$ to $L^k$, and update the current sequence position: $i \gets i+1$.
\item Otherwise we have a match $S^k_{i,j} = R_{p, p+j-i}$ of length $j-i+1$.
We encode it by appending the tuple $\langle f_\text{match\_1st\_lev}, p,\allowbreak j-i+1\rangle$ to $L^k$.
Then, we update the current sequence position: $i \gets j+1$.
\end{itemize}
%The tuple flags $f_\text{literal}, f_\text{match\_1st\_lev}$ distinguish between a literal and match tuples in a sequence of tuples, which is usually implemented in the Ziv-Lempel algorithms as a sequence of integers.

There is, however, some exception to the general rule that no shorter than $h_{1m}$ symbols match can be found.
Genomic sequences often differ by single nucleotide polymorphism (SNPs) or short indels (a few symbols long insertions or deletions).
Thus, when some match is found, before looking for another match in~$R$ using the hash table $HT^\text{R}$, we do 3 (or 5, depending on the user-specified option) simple verifications.
We check whether the next symbol(s) after the current match is just a single nucleotide mutation or a single-symbol (or double-symbol) indel.
We allow matches found after such variation to be of length $h_{1e}$ (equal to 4 by default).
The rationale for such decision is two-fold.
Firstly, it speeds up the searching as for the verification we do not need to query the hash table $HT^\text{R}$.
Secondly, such matches (even if they are short) can be quite efficiently encoded as the match position is easy to predict (encoding of Ziv--Lempel parsing results is described below). 
Thus, even if the sequence~$L^k$ will be longer when such short matches are allowed, the final compression ratio can be better.

At the second level, the algorithm performs a similar Ziv--Lempel factoring of the collection $\mathcal{L} = \{L^1, L^2, \ldots, L^n\}$ to obtain the collection $\mathcal{D} = \{D^1, D^2, \ldots, D^n\}$.
We will use here similar notations as for the sequences~$\mathcal{S}$, i.e., $l^k_i$ is the $i$th tuple of sequence~$L^k$, $L^k_{i,j}$ is $l^k_i l^k_{i+1} \ldots l^k_j$.
Additionally we define the \emph{weight} of a substring $L^k_{i,j}$ as the sum of weights of the tuples it is composed of, where the weight of a literal tuple is~1 and the weight of a match tuple is~7 (values chosen experimentally).
A~search structure $\mathit{HT}^\text{L}$ (namely, hash table with linear probing) is used here to look for matches in $\mathcal{L}$.
At the beginning $\mathit{HT}^\text{L}$ is empty, but we update it by adding the already processed sequences of $\mathcal{L}$, i.e., when processing $L^k$ the hash table $\mathit{HT}^\text{L}$ contains all substrings of tuples of weights ``close'' to $h_2=11$
%\footnote{For each position $i$ in the tuple sequence $L^u$ we take the shortest substring (in terms of the number of tuples) $L^u_{i,j}$ of weight not smaller than $h_2$.} 
of $L^1$, $L^2$, \ldots, $L^{k-1}$.
(For each position $i$ in the tuple sequence~$L^u$ we take the shortest substring (in terms of the number of tuples) $L^u_{i,j}$ of weight not smaller than $h_2$.)

Now, when we process $L^k$ starting from $i=1$ to obtain $D^k$, we look for the match of the largest weight $L^k_{i,j} = L^u_{p, p+j-i}$.
There are two possible situations here:
\begin{itemize}
\item No match of weight at least $h_2$ is found.
In this case we append the tuple~$l^k_i$ (describing the first level literal or the first level match) to $D^k$ and update the current sequence position: $i \gets i+1$.
\item Match $L^k_{i,j} = L^u_{p, p+j-i}$ is found.
In this case we append the tuple $\langle f_\text{match\_2nd\_lev}, u, i, j-i+1\rangle$ to $D^k$ and update the current sequence position: $i \gets j+1$.
\end{itemize}

The sequence $D^k$ is composed of tuples of three kinds: first level literal (pair), first level match (triple), second level match (quadruple).
%The flag $f_\text{match\_2nd\_lev}$ in the quadruple is to distinguish the type of tuple in real implementations, when they are stored as sequence of integers.
Since when processing~$L^1$ the search structure $\mathit{HT}^\text{L}$ is empty, $D^1 = L^1$.

\begin{figure}[ht]
\center
%\footnotesize
%\begin{verbatim}
%     12345678901234567890
% R = ACTGACCGTCGATTTAACCC
%
%S1 = AACTGACCGTAAGATTTAATGC
%S2 = ACTAACCGTCCATTTAATGC  
%S3 = ACTAACCGTCAATTTAATGC
%S4 = AAACTGATCGTCGTTAAGTGC
%S5 = ACTGATCGTCGTTAACCC
%S6 = ACTGATCGTCGTTAAGCC
%
%
%h1=3, h1'=2 (after snp/del/ins), h2=3,
%weights: L1M=2 (level_1_match), L1L=1 (level_1_literal)
%
%
%L1 = (L1M, 16, 3), (L1M, 3, 7), (L1L, A), (L1L, A), (L1M, 11, 7), (L1L, T), (L1L, G), (L1L, C)
%
%L2 = (L1M, 1, 3), (L1L, A), (L1M, 5, 6), (L1L, C), (L1M, 12, 6), (L1L, T), (L1L, G), (L1L, C) 
%
%L3 = (L1M, 1, 3), (L1L, A), (L1M, 5, 6), (L1L, A), (L1M, 12, 6), (L1L, T), (L1L, G), (L1L, C) 
%
%L4 = (L1L, A), (L1M, 16, 3), (L1M, 3, 3), (L1L, T), (L1M, 7, 5), (L1M, 13, 2), (L1M, 16, 2), (L1L, G), 
%     (L1L, T), (L1L, G), (L1L, C) 
%
%L5 = (L1M, 1, 5), (L1L, T), (L1M, 7, 5), (L1M, 13, 2), (L1M, 16, 5)
%
%L6 = (L1M, 1, 5), (L1L, T), (L1M, 7, 5), (L1M, 13, 2), (L1M, 16, 2), (L1L, G), (L1M, 19, 2) 
%
%
%D1 = (L1M, 16, 3), (L1M, 3, 7), (L1L, A), (L1L, A), (L1M, 11, 7), (L1L, T), (L1L, G), (L1L, C) 
%
%D2 = (L1M, 1, 3), (L1L, A), (L1M, 5, 6), (L1L, C), (L1M, 12, 6), (L2M, L1, 6, 3)
%
%D3 = (L2M, L2, 1, 3), (L1L, A), (L2M, L2, 5, 4) 
%
%D4 = (L1L, A), (L1M, 16, 3), (L1M, 3, 3), (L1L, T), (L1M, 7, 5), (L1M, 13, 2), (L1M, 16, 2), (L1L, G), 
%     (L2M, L1, 6, 3)  
%
%D5 = (L1M, 1, 5), (L2M, L4, 4, 3) (L1M, 16, 5)
%
%D6 = (L2M, L5, 1, 4), (L2M, L4, 7, 2), (L1M, 19, 2) 
%\end{verbatim}
\includegraphics[width=\textwidth]{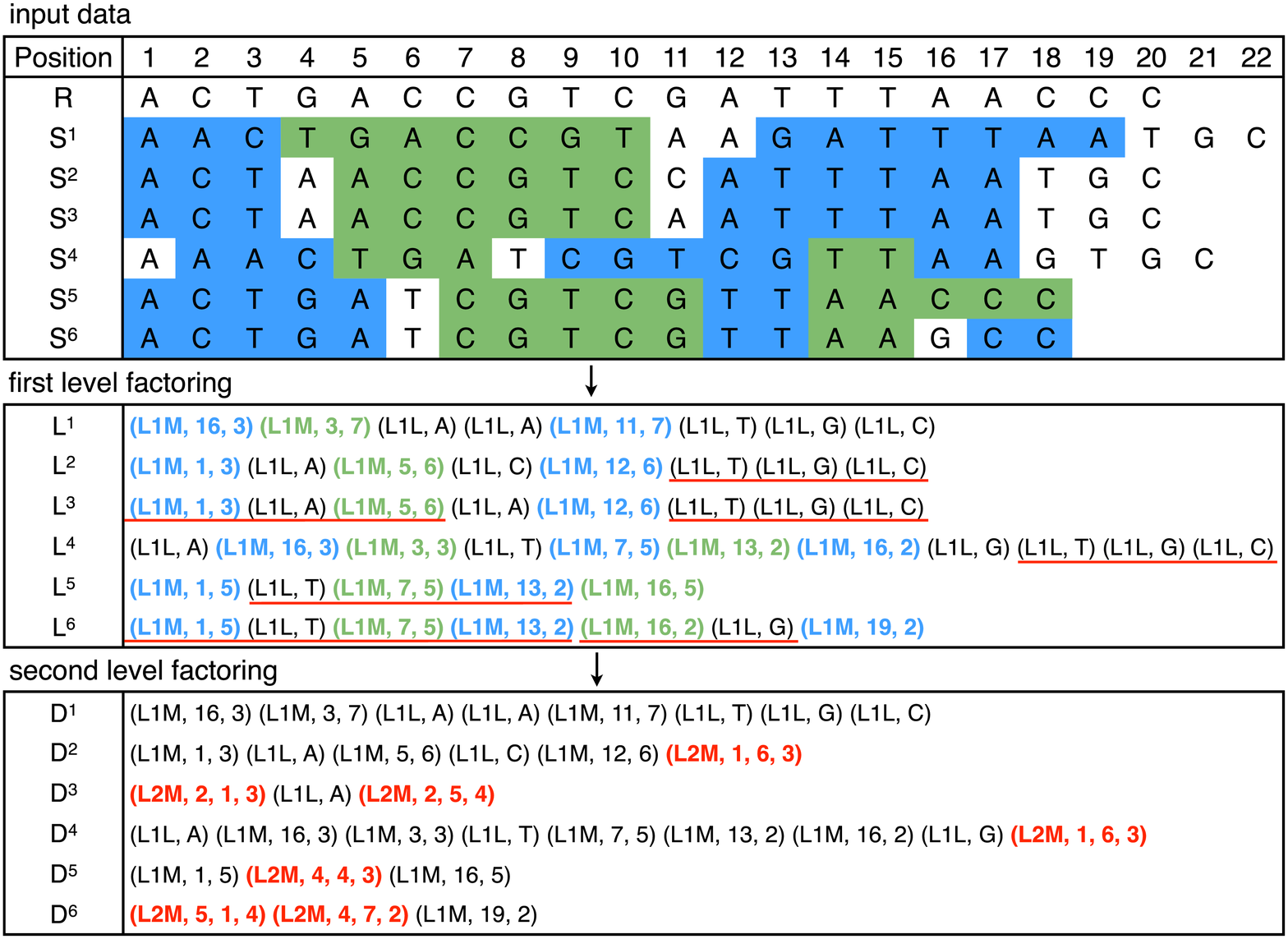}
\caption{Example of first and second level factoring in GDC 2 algorithm, where: $h_{1m}=3$, $h_{1e}=2$, $h_{2}=3$, weight of a literal tuple is $1$ and weight of a match tuple is 2. 
Blue and green colors are used only to distinguish between adjacent first-level matches.
The red underline is to point the second-level matches.
The used abbreviations: L1L --- $f_\text{literal}$, L1M --- $f_\text{match\_1st\_lev}$, L2M --- $f_\text{match\_2nd\_lev}$.
}
\label{fig:LZ-factoring}
\end{figure}

The reason for using two-level Ziv--Lempel factoring is that the genome sequences are usually highly similar, so in the whole collection the same series of matches and literals between the current sequence and the reference sequence can be found.
Thus, instead of storing the series of tuples many times, it is beneficial to encode them once and only reference to them for other sequences.
Figure~\ref{fig:LZ-factoring} shows how the two-level factoring is performed.

The collection $\mathcal{D}$ is a succinct representation of the input collection $\mathcal{S}$.
Nevertheless, it has potential to be compressed even more if we use an arithmetic coder.~\cite{SM2010}
What is important, instead of encoding the tuples as they are, we predict some of their values (e.g., matching positions) and encode only the differences between our predictions and the real values.
The successive fields of the tuples are arithmetically encoded as follows.

\subsubsection*{Flags} 
There are only 3 different flags distinguishing between the tuple types.
We encode them contextually, where the context is composed of two recently encoded flags.

\subsubsection*{Codes of symbols in the first level literals} 
Codes of symbols are encoded contextually, where the context is the recently encoded symbol.

\subsubsection*{Positions of the first level matches}
These positions can be from a broad range, i.e., between $1$ and almost $|R|$.
Since, the genomic sequences are similar, the position of the current match is likely to be close to the position of the previous match increased by the number of symbols encoded in the meantime.
Thus, before encoding the position $\mathit{pos}$ we estimate its value $\mathit{expected\_pos}$ and encode only the difference $\mathit{relative\_pos} = \mathit{expected\_pos} - \mathit{pos}$.
The $\mathit{expected\_pos}$ is calculated by increasing the recently encoded $\mathit{pos}$ by: (\emph{i})~the length of the last match, (\emph{ii}) the number of literals encoded since the last match, (\emph{iii}) the number of symbols encoded as the second level matches seen from the recent  first level match.
Then, the estimation is classified as: \emph{perfect} ($\mathit{relative\_pos} = 0$), \emph{good} ($0 < |\mathit{relative\_pos}| < 2^6$), \emph{poor} (other values).
Finally, the estimation type is encoded without a context and the necessary number of bytes (0, 1, or 4) of $\mathit{relative\_pos}$ are encoded with context being the estimation type and number of encoded byte.

\subsubsection*{Lengths of the first level matches}
Each length is classified as: \emph{short} (not longer than $2^8$ symbols), \emph{long} (of length between $2^8$ and $2^{16}+2^8$ symbols), \emph{very long} (longer than $2^{16}+2^8$ symbols).
Then, the length type is encoded (without a context).
Finally, the necessary number of bytes (1, 2, or 4) of the length are encoded with context being the length type and the number of encoded byte.

\subsubsection*{Sequence ids of the second level matches}
The value $\mathit{id}$ is split into two integers: $\lceil\mathit{id} / 256\rceil$ (prefix) and $\mathit{id} - 256 \times \lceil\mathit{id} / 256\rceil$ (suffix).
The prefix is encoded without a context.
The context of the suffix is the prefix.

\subsubsection*{Positions of the second level matches}
Similarly like the positions of the first level matches, these values can be from a broad range.
Thus, instead of encoding them as they are, we estimate the position and encode only the difference.
Let us assume the current sequence is $L^k$.
We need some auxiliary array $A[1..k]$ to make the estimations possible. 
Now we will discuss how $A$ is maintained when processing~$L^k$.
Then, we will show how it is used to estimate the positions of the second level matches.

Let us assume that the we have a match in the sequence $L^u$.
After encoding it we store in $A[u]$ the pair $\langle p^\text{A}, s^\text{A}\rangle$, where  $p^\text{A}$ is the match position in $L^u$ and $s^\text{A}$ is the number of symbols of $S^k$ processed before the current match.

Thus, the encoding of the match positions is made as follows.
For a match in the sequence~$L^u$ we calculate the difference~$d$ between the current position in $S^k$ and the position $s^\text{A}$ stored in $A[u]$.
Then, we advance the position $p^\text{A}$ (stored in $A[u]$) of $L^u$ as long as the number of the symbols covered by the first level literals and matches is not larger than~$d$.
What we obtain is the expected position in $L^u$ for the current match.

Then, we can calculate the difference between the expectation and the value of the current tuple.
The estimations are classified as: \emph{perfect} (difference is 0), \emph{good} (absolute value of the difference between $1$ and $16$), \emph{moderate} (absolute value of the difference between $16$ and $256$), and \emph{poor} (other values).
Finally, the estimation type is encoded without a context and the necessary number of bytes of the difference are encoded with context being the estimation type and the number of the encoded byte.

\subsubsection*{Lengths of the second level matches}
The lengths are classified according the their value to: \emph{short} (not longer than $2^4$ tuples), \emph{medium-sized} (between $2^4$ and $2^5+2^4$ tuples), \emph{long} (between $2^5+2^4$ and $2^7+2^5+2^4$ tuples), \emph{very long} (between $2^7+2^5+2^4$ and $2^8+2^7+2^5+2^4$ tuples), \emph{extremely long} (the rest).
Then, the length type is encoded (without a context).
Finally, the necessary number of bytes is encoded, where the context is the length type and additionally (for extremely long lengths) also the number of encoded byte.

\subsection*{Decompression algorithm}
Decompression is straightforward.
At the beginning the $\mathcal{D}$ collection is obtained by arithmetically decoding the compressed file.
Then, the collection $\mathcal{L}$ is decoded.
Finally, the sequences of $\mathcal{S}$ are constructed from $\mathcal{L}$ and~$R$.

\subsection*{Access to a single compressed sequence}
A drawback of the proposed algorithm is that to decompress $S^n$ we need to decompress (at least at the second level) all other sequences.
More precisely, to obtain $S^m$ we need to have $L^1, L^2,\allowbreak{} \ldots, L^{m-1}$ as they must be known to obtain $L^m$.
Then, we can obtain $S^m$ from $L^m$ and $R$.
This can be important especially when $m$ is large.
To partially solve this problem we implemented a variant of the compression algorithm in which we allow to set by the user (during compression) the fraction of the sequences that can be used as the second-level references.
Thus, when this parameter is, e.g., 30\%, in the worst case only 30\% of $\mathcal{L}$ must be decompressed.
This deteriorates the compression ratio, so this is rather a compromise than a perfect solution.

\subsection*{Real implementation}
To increase the speed of the compression and decompression we designed the compressor in a multithreaded fashion.
There are several (user-defined) threads performing the first level compression (and decompression) and a single thread performing the second level compression (and decompression).
For example, in the compression, each of the first level threads reads a sequence $S^k$ from a queue of sequences to compress and performs the Ziv--Lempel factoring of $S^k$ according to~$R$.
The results~$L^k$ are stored in an in-memory queue $Q$.
The second level compression thread reads sequences $L^k$ from $Q$, performs the Ziv--Lempel factoring of it according to the already processed part of sequences from $\mathcal{L}$ obtaining $D^k$ and finally performs also the entropy coding of $D^k$.
(We use a popular and fast arithmetic coding variant by Schindler, also known as a range coder (\url{http://www.compressconsult.com/rangecoder/}).)
The queue $Q$ has FIFO (first in first out) organization, so there is no guarantee in which order the sequences of $\mathcal{L}$ will be processed (it depends on the processing time of the sequences by the first level threads).
Thus, the compression ratios can slightly differ between the executions of the algorithm.

The parallel design of the decompression algorithm is similar.

The compression output is composed of three files.
The one with extension \textsf{gdc2\_desc} stores file names, sequence sizes, and ids of the multi-FASTA sequences.
It is small, but to provide the best possible compression ratio of the whole algorithm, it is compressed using popular zlib library.
The file with extension \textsf{gdc2\_rc} contains the compressed representation of the collection~$\mathcal{S}$.
Finally, the file with extension \textsf{gdc2\_ref} stores the compressed reference sequence~$R$.
As it is not a part of the collection to be compressed, its size is not counted in the experimental results.
Nevertheless, we decided to compress it for the situations in which the user is interested in storing both the reference $R$ and the collection~$\mathcal{S}$ in a single place in a compact form.
This file is compressed by gathering symbols in triples and encoding them arithmetically.

%
%\begin{figure}
%...
%\caption{Basic scheme of the compression algorithm}
%\label{fig:scheme}
%\end{figure}

\subsection*{Relation of the proposed compressor to the existing works}
The proposed compressor bares some similarities to the existing works.
The main concept of two-level Ziv--Lempel factoring is an extension of what was done in FRESCO~\cite{WL2013}.
In FRESCO, the collection of sequences is split into two sets: additional references and the remaining sequences.
The additional references are compressed only according to the main reference sequence.
The remaining sequences are compressed only according to the main and the additional references.
In GDC~2, we do not split the collection into two sets.
We just use all of the already processed sequences as the additional references for the current sequence, with significant boost in the compression ratio.
%The efficient encoding of the sequence of tuples, especially, the prediction of positions of matches has nothing common with FRESCO.
Moreover, FRESCO uses LZ77 factoring~\cite{ZL1977}, while GDC~2 uses LZSS factoring.~\cite{SS1982}

The concept of looking for short matches after some longer ones is an extension of what was made in our previous work~\cite{DG2011}.
In GDC~2 we, however, allow not only single-letter mismatches, but also short indels.
We also do not limit the number of short matches in a series.
The way the tuples are encoded using an arithmetic coder, especially the calculation of the expected positions for the first- and second-level matches, is novel in this context.

Also the multithreaded design of GDC~2 was not used by existing multi reference genome compressors.

\section*{Results}
Our compressor, GDC~2, was implemented in C++11 language using C++ built-in concurrency mechanisms.
The test machine was equipped with Intel i7 4930K CPU (6 cores, clocked at 3.4\,GHz), 64\,GB of RAM,  and two 3\,TB HDDs in RAID 0 (measured average read speed about 350\,MB/s).

For the experiments we used two large datasets.
% (see Table~\ref{tab:datasets} for characteristics).
%They were formerly used in various papers on genomic data compression.
\emph{A.thaliana} dataset of total size 94\,GB was obtained from the 1001GP~\cite{1001GP} and contains 775 sequences.
%It contains longer sequences 
\emph{H.sapiens} dataset of total size 6670\,GB was obtained from the 1000GP~\cite{TGP2012} and contains 2184 sequences (from 1092 diploid human genomes).

%\begin{table}[ht]
%\begin{tabular}{lccc}\toprule
%Dataset			& No.\ of sequences	& Ref.\ genome length [Mbp]	& Total size [GB] \\ \midrule
%?? S.cerevisiae	& \qq38		\\
%?? S.paradoxus		& \qq35		\\
%A.thaliana		& \q775					& 	\qc120	& \qqc94.0\\
%H.sapiens		& 2184					&	3,093	& 6,669.8\\
%\bottomrule
%\end{tabular}
%\caption{Characteristics of the test datasets.
%The no.\ of sequences does not include the reference sequence}
%\label{tab:datasets}
%\end{table}

The comparison of all of the existing genomic data compressors would be very hard due to many problems.
For example, some compressors do not support symbols other than \textsf{ACGT}, some cannot work with so huge data, some are very slow and performing complete experiments would take months.
Thus we selected the compressors that proved to be the best (in terms of compression ratio) in the previous studies:
7z (general purpose compressor from the Ziv--Lempel family), RLZ~\cite{KPZ2011}, GReEn~\cite{PPG2012},  ABRC~\cite{WL2012}, GDC normal~\cite{DG2011}, GDC ultra~\cite{DG2011}, iDoComp~\cite{OHW2014}, FRESCO~\cite{WL2013}.f
In the preliminary experiments (Table~\ref{tab:preliminary}), we evaluated them on subsets of our datasets to select the candidates for more complete evaluation.
As the results show, the single-reference compressors (RLZ, GReEn, ABRC, GDC-normal, iDoComp) give ratios much smaller than 1000 for \emph{H.sapiens} chromosomes and smaller than 160 for \emph{A.thaliana} chromosomes.

\begin{table}[ht]
\centering{\small
\begin{tabular}{lcccccccc}
\toprule
Dataset			& 7z		& RLZ	& GReEn	& ABRC	& GDC-normal 	& iDoComp	& GDC-ultra 	& FRESCO	\\
%					& 			& 		& 			& 			& normal	& 				& ultra	&			\\
\midrule
\multicolumn{9}{c}{\emph{\bfseries H.sapiens}}\\
Chr.\ 14			& 1,068	& 270	& 218		& 472		& 674		& 625		& 2,455	& 1,946		\\
Chr.\ 21			& 1,561	& 269	& 211		& 460		& 685		& 642		& 2,397	& 	2,545		\\
\midrule
\multicolumn{9}{c}{\emph{\bfseries A.thaliana}}\\
Chr.\ 1	& \qc242	&\q86	&\q64		&\q67		& 154				&	156	&\qc254	& \qc186	 \\
Chr.\ 4	& \qc234	&\q80	&\q59		&\q61		& 141				&	145	&\qc230	& \qc170	 \\
\bottomrule
\end{tabular}}
\caption{Compression ratios for subsets of the datasets for various compressors}
\label{tab:preliminary}
\end{table}

The general purpose 7z can be seen as a multi-reference compressor since it looks for matches between the present sequence and the sequences seen in the past 1\,GB.
For \emph{H.sapiens} Chromosome 21 it means about 20 recently processed sequences.
Nevertheless, for \emph{H.sapiens} Chromosome 1 these would be only 4 sequences.
The true multi-reference compressors GDC-ultra and FRESCO give much better ratios for human chromosomes.
For FRESCO we set the number of additional reference sequences to 100 as in a preliminary experiment (results not shown) this leaded to better compression ratios than the value 70 used in the original paper~\cite{WL2013}.

In a consequence, for further experiments we selected two best single-reference compressors, i.e., GDC-normal and iDoComp, and two best multi-reference compressors, i.e., GDC-ultra and FRESCO.
The results of evaluation of the chosen compressors and the proposed GDC~2 are presented in Tables~\ref{tab:h.sapiens} and~\ref{tab:a.thaliana}.
For the \emph{H.sapiens} dataset (Table~\ref{tab:h.sapiens}) the compression ratio of GDC~2 is about 9500, which is approximately 4 times better than the best of the existing competitors.

\begin{table}[htp]
\centering{\small
\begin{tabular}{lcccccc}
\toprule
Data			&	Raw size	& GDC normal& iDoComp	& GDC ultra & FRESCO	& GDC 2\\
				& [GB]		& ratio		& ratio		& ratio		& ratio	& ratio\\
\midrule
Chr.\ 1		&\qc551.7\q	& \qqc680	& \qqc659	&\q2,508		&\q2,279 & \bf\q10,556	\\
Chr.\ 2		&\qc 538.5\q& \qqc628	& \qqc608	&\q2,318	 	&\q2,113	& \bf\qq9,828\\
Chr.\ 3		&\qc 438.4\q& \qqc602	& \qqc552	&\q2,263		&\q2,044	& \bf\qq9,564\\
Chr.\ 4 		&\qc 422.8\q& \qqc547	& \qqc503	&\q2,202		&\q1,911	& \bf\qq8,979\\
Chr.\ 5		&\qc 400.6\q& \qqc624	& \qqc576	&\q2,260		&\q1,997	& \bf\qq9,578\\
Chr.\ 6		&\qc 378.7\q& \qqc566	& \qqc522	&\q2,184		&\q1,950	& \bf\qq8,832\\
Chr.\ 7		&\qc 352.3\q& \qqc592	& \qqc545	&\q2,138		&\q1,918	& \bf\qq8,752\\
Chr.\ 8		&\qc 323.9\q& \qqc584	& \qqc543	&\q2,137		&\q1,916	& \bf\qq8,817\\
Chr.\ 9		&\qc 312.5\q& \qqc718	& \qqc666	&\q2,450		&\q2,359	& \bf\q10,400	\\
Chr.\ 10		&\qc 300.1\q& \qqc578	& \qqc564	&\q2,123		&\q1,973	& \bf\qq9,335\\
Chr.\ 11		&\qc 298.8\q& \qqc560	& \qqc521	&\q2,171		&\q1,967	& \bf\qq9,043\\
Chr.\ 12		&\qc 296.2\q& \qqc595	& \qqc547	&\q2,167		&\q1,958	& \bf\qq9,127\\
Chr.\ 13		&\qc 255.0\q& \qqc611	& \qqc564	&\q2,452		&\q1,842	& \bf\q10,669	\\
Chr.\ 14		&\qc 237.6\q& \qqc674	& \qqc625	&\q2,458		&\q1,946	& \bf\q10,654	\\
Chr.\ 15		&\qc 227.0\q& \qqc716	& \qqc664	&\q2,458		&\q2,020	& \bf\q10,815	\\
Chr.\ 16		&\qc 200.1\q& \qqc647	& \qqc604	&\q2,068		&\q2,076	& \bf\qq8,980\\
Chr.\ 17		&\qc 179.7\q& \qqc646	& \qqc594	&\q2,059		&\q2,090	& \bf\qq8,651\\
Chr.\ 18		&\qc 172.9\q& \qqc568	& \qqc525	&\q2,051		&\q2,066	& \bf\qq9,033\\
Chr.\ 19		&\qc 130.8\q& \qqc569	& \qqc519	&\q1,773		&\q1,828	& \bf\qq7,137\\
Chr.\ 20		&\qc 139.5\q& \qqc633	& \qqc619	&\q2,014		&\q2,240	& \bf\qq9,150\\
Chr.\ 21		&\qc 106.5\q& \qqc686	& \qqc642	&\q2,405		&\q2,545	& \bf\q10,414	\\
Chr.\ 22		&\qc 113.5\q& \qqc823	& \qqc772	&\q2,455		&\q2,718	& \bf\q10,547	\\
Chr.\ X-fem	&\qc 178.5\q& \qqc911	& \qqc826	&\q2,551		&\q2,628	& \bf\q11,060	\\
Chr.\ X-mal	&\qqc81.0\q	& \qqc945	& \qqc896	&\q2,740		&\q2,469	& \bf\q11,546	\\
Chr.\ Y-mal	&\qqc30.0\q	& 38,233		& 59,062		& 42,870		& 39,228	& \bf132,123\\
Chr.\ X-mal1&\qc\qq2.8\q& \qqc312	& \qqc310	&\qqc587		&\qqc713	& \bf\qq2,423\\
Chr.\ X-mal2&\qc\qq0.35	& \qqc280	& \qqc456	&\qqc741		&\qqc943	& \bf\qq5,914\\
\midrule
Complete		&6,669.8\q&\qqc627	& \qqc586	&\q2,262		&\q2,065	&\bf\qq9,557 \\
\midrule
Compression speed [MB/s]	&				&	\qqc\q73		& \qqc\q51	&	\qqc\q12	&\qqc111	&\bf\qqc\q 202	 \\
%D.\ speed  \\
%D.\ speed w/o I/O	&& \\
\bottomrule
\end{tabular}}
\caption{Compression ratios for \emph{H.sapiens} dataset.
The ratios are calculated as raw size divided by compressed size rounded to the integer.
Compression speeds (in MB/s) are given in the bottom line of the table.
%The `w/o I/O' speed is when the output of the decompressor is sent to /dev/null.
Raw sizes are in GBs.}
\label{tab:h.sapiens}
\end{table}

\begin{table}[htp]
\centering{\small
\begin{tabular}{lcccccc}
\toprule
Data			&	Raw size	& GDC normal& iDoComp	& GDC ultra & FRESCO	& GDC 2\\
				& [GB]		& ratio		& ratio		& ratio		& ratio	& ratio\\
\midrule
Chr.\ 1		& 23.9\q		& 154			& \qc156		& 254			& \qc186	&\bf\qqc621	\\
Chr.\ 2		& 15.5\q		& 143			& \qc148		& 239			& \qc175	&\bf\qqc559 \\
Chr.\ 3		& 18.4\q		& 147			& \qc152		& 238			& \qc169	&\bf\qqc551 \\
Chr.\ 4 		& 14.6\q		& 141			& \qc145		& 230			& \qc170	&\bf\qqc553 \\
Chr.\ 5		& 21.2\q		& 148			& \qc151		& 254			& \qc187	&\bf\qqc624 \\
Chr.\ C		& \q0.12		& 652			&	1,830		& 652			& 1,750	&\bf25,061	\\
Chr.\ M		& \q0.29		& 558			& \qc807		& 600			& \qc374	&\bf\q1,401	\\
\midrule
Complete		& 94.0\q		& 148			&\qc151		& 245			& \qc179	&\bf\qqc587 \\
\midrule
Compression speed [MB/s] &				& 120			& \qqc47		& \q13		& \qqc\q7& \bf\qqc\q94 \\
%D.\ speed  \\
%D.\ speed w/o I/O	&& \\
\bottomrule
\end{tabular}}
\caption{Compression ratios for \emph{A.thaliana dataset}.
The ratios are calculated as raw size divided by compressed size rounded to the integer.
Compression speeds (in MB/s) are given in the bottom line of the table.
%The `w/o I/O' speed is when the output of the decompressor is sent to /dev/null.
Raw sizes are in GBs.}
\label{tab:a.thaliana}
\end{table}

%The bottom lines of the tables show the compression and decompression speeds.
In the compression, the fastest is GDC~2, which works with a speed about 200\,MB/s.
Measuring of the speed of decompression is problematic as some of the compressors work faster than the disk speed ($\sim$350\,MB/s), which in practice is more than sufficient.
Nevertheless, we were interested in what is the true decompression speed of the GDC~2 algorithm, so we measured it with the output redirected to /dev/null (i.e., the sequences were decompressed but not stored) obtaining about 1000\,MB/s.
%In practice, speeds better than disk speed are more than sufficient, but we were also interested in what is the true speed of the algorithm. 
%Therefore, we measured the speeds of the programs where the output was redirected to /dev/null. 
%As we can see the speed of GDC~2 is about 1300\,MB/s.

The experiment for the \emph{A.thaliana} dataset (Table~\ref{tab:a.thaliana}) shows that the compression ratios are much worse.
The best ratio, almost 600 was obtained by GDC~2.
This result is approximately 2.4 times better than the second best, GDC-ultra.
Also the compression speeds are worse here.

\begin{figure}[t]
\includegraphics[width=\textwidth]{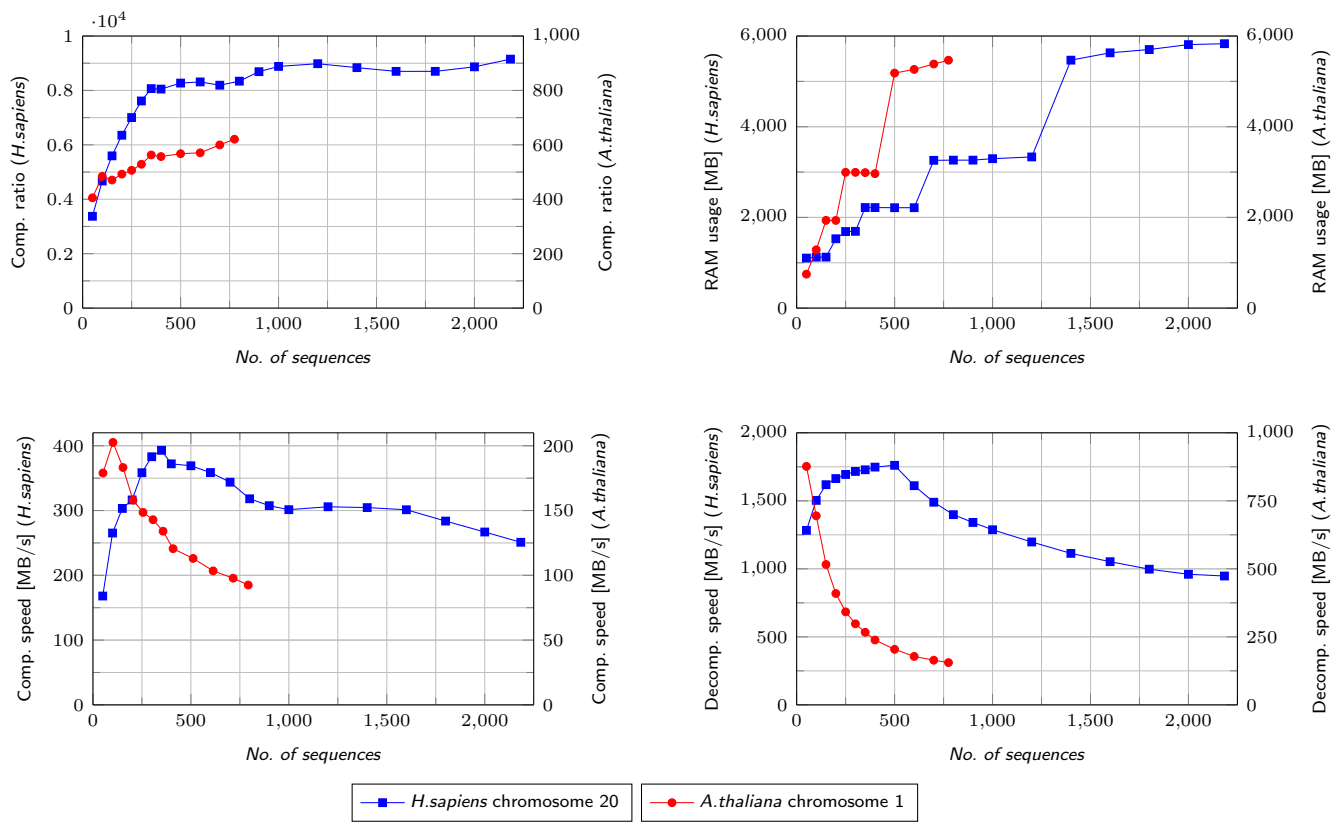}
\caption{Influence of the number of sequences in the input collection on: compression ratio (left top), memory usage (right top), compression speed (left bottom), decompression speed (right bottom).
The decompression speed was measured when the output was redirected to /dev/null, i.e., the sequences were decompressed but not stored.}
\label{fig:chart_seqs}
\end{figure}

In the next experiment, we measured the influence of the number of sequences in the input collection on the compression ratio, compression and decompression speeds, and memory usage.
The results for two chromosomes are shown in Figure~\ref{fig:chart_seqs}.
As one can see for the human chromosome the compression ratio rapidly achieves about 8000 for 300 input sequences and then grows moderately.
The same phenomenon can be observed for {\em A.thaliana} data, but the ratio is about an order of magnitude lower.

The memory usage of GDC~2 depends mainly on the number of sequences serving as the second level references as they must be stored (and indexed) in memory during compression.
In this experiment all sequences were used as additional references, so the memory consumption grew constantly up to about 6\,GB.
(The most memory consuming was compression of {\em H.sapiens} Chromosome 2 for which about 24\,GB of RAM was necessary.) 
The visible stepwise increment of the memory usage is a consequence of the assumed possible hash table size (being always a power of~2).

The compression and decompression speeds for the human dataset initially grow with the increasing number of sequences and are the highest for the collection of size about 300--500. This is correlated with the growing compression ratio. 
Roughly speaking, the more second level references, the better the second-level factoring (i.e., longer matches can be found) and so, there are significantly less data to process by the arithmetic coder. 
However, for larger collections, much more data must be analyzed during the second level factoring, so the speed of compression falls down.
A similar thing happens in the decompression.
The better second-level factoring means less data to be arithmetically decoded, which increases the speed.
Unfortunately, more second-level references means much more computations for the estimation of the positions of matches and this term dominates for large collections.
%\new{The decompression speed  ...}

In the next experiment, we measured the influence of the number of reference sequences in the second level of GDC 2 on the compression ratio,  (de)compression speeds and the extraction time of a single sequence of a collection. 
The most important results are presented in Figure~\ref{fig:chart_percent} (the complete results are in Supplementary Figure S1). 
Decreasing the number of second level references by half results in a reduced RAM usage (about half less RAM is used) and a noticeable speed up of compression (24\% for \emph{H.sapiens} dataset and 17\% for \emph{A.thaliana} dataset) at a cost of some decrease of compression ratio (26\% and 14\%, respectively). Using even less sequences in the second level of GDC 2 leads also to significant gains in speed of decompression of complete collection or a single sequence, obviously at a cost of decreased compression ratio. For 10\% of the sequences used, average single sequence access times decreased from 53 to 31 seconds for \emph{H.sapiens} dataset (at a cost of 2.85 worse compression ratio) and from 63 to 21 seconds for \emph{A.thaliana} dataset (at a cost of 1.79 worse compression ratio). 

% For 10% of the sequences used, single sequence access time is 1.7 times better for H.sapiens data and more than 3 times better for A.thaliana data, while the compression ratios are 2.85 and 1.79 times worse, respectively.

\begin{figure}[ht]
\includegraphics[width=\textwidth]{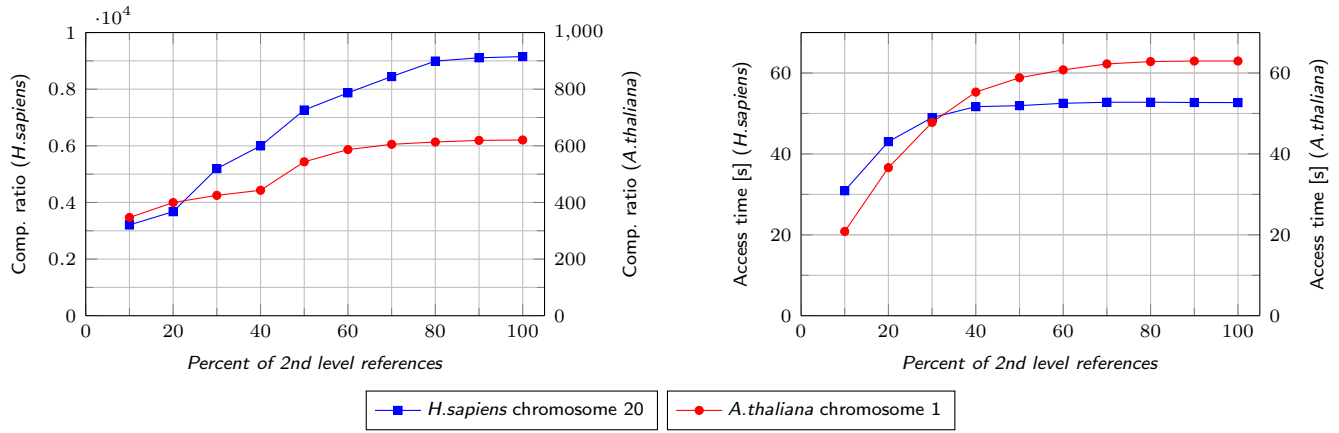}
\caption{Influence of the percent of 2nd level references on compression ratio (left), decompression (access) time of a single sequence (right).}
\label{fig:chart_percent}
\end{figure}

\begin{figure}[ht]
\includegraphics[width=\textwidth]{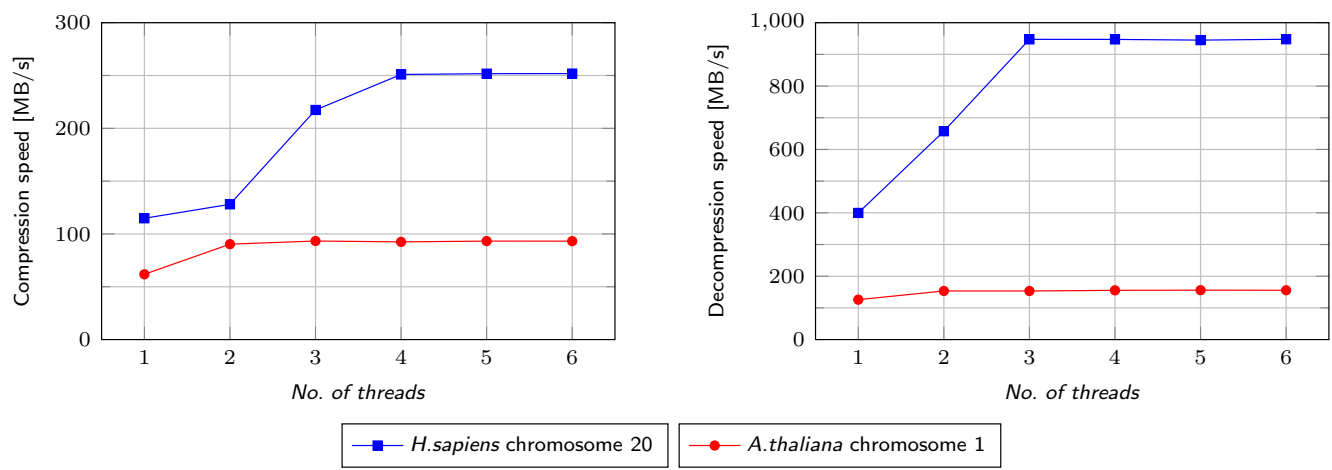}
\caption{Influence of the number of threads used by GDC~2 algorithm on compression and decompression speeds.}
\label{fig:chart_threads}
\end{figure}

GDC~2 is implemented in a multithreaded fashion, so it is natural to ask how its speed scales when when the number of threads is increased.
By default, GDC~2 uses 4 threads: 3 for the first level Ziv--Lempel factoring and 1 for the second level factoring and arithmetic coding.
The results presented in Figure~\ref{fig:chart_threads} show that the value 3 or 4 seems to be an optimal choice.
The speed is limited by disk speed or (for fast disks) by the single second level compressing thread.
This suggest that splitting this thread into two, e.g., one performing Ziv--Lempel factoring and other performing arithmetic compression would increase the total performance of GDC~2.
Nevertheless, since the absolute values of compression speeds are high, we resigned from that in the present version of the software.

\section*{Discussion}
We proposed the new algorithm for compression of collections of complete genome sequences.
% genomic sequence collections.
The evaluation shows that its compression ratios are roughly 4 times better than the best existing competitors.
Moreover, it is very fast, as the compression speed for the human data set is about 200\,MB/s.
The decompression speed is limited by the speed of the disk used in the experiments.
When we measured this speed without storing the files onto disks, it was about 1000\,MB/s.
The algorithm is designed primarily to compress and decompress efficiently a large collection of genomes all at once. However, extraction of a single sequence is also possible. The access time, although not impressive (counted in tens of seconds), can be significantly improved at a cost of some decrease in an overall compression ratio.

It is also interesting to compare the compression ratios with what is possible, when much more knowledge of the data is given.
Namely, when the input data are given as differences between the sequences and the reference (in VCF format), the best compressor, TGC, was able to obtain even better ratios.
For human data set they are about 15,500.
When we compare this with about 9,500 of GDC~2 we see that we are quite close to what is theoretically possible.
Similar results are for \emph{A.thaliana} dataset: $\sim$590 ratio for GDC~2 and $\sim$860 ratio for TGC.
What is, however, worth to stress, GDC~2 is able to compress collections of sequences of the same species gathered from various sources (e.g., {\em de novo} assembled), when no alignment of them is given, while TGC input must be perfectly aligned sequences described as variants between them.

%\section*{References}

%Books:
%
%    Smith, J. Syntax of referencing in How to reference books (ed. Smith, S.) 180-181 (Macmillan, 2013).
%    Jones, H. in How to reference books 2nd edn, Vol. 4 (eds Brown, R. et al.) Ch. 8, 222-223 (Macmillan, 2014).
%
%Website:
%Buckley, M. and Reid, A., Global food safety - keeping food safe from farm to table. Technical report. (2010) Available at: insert website here. (Accessed: 4th November 2012)
%

\section*{Acknowledgments}
The Polish National Science Centre under the project DEC-2011/03/B/ST6/01588.
The infrastructure supported by POIG.02.03.01-24-099/13 grant: `GeCONiI---Upper
Silesian Center for Computational Science and Engineering'.

\section*{Author contributions statement}
SD and AD designed the algorithm.
SD, AD, and MN prepared the implementation.
SD and AD performed the experiments.
SD and AD drafted the manuscript and the supplementary material.
All authors read and approved the final manuscript.

\section*{Additional information}
The supplementary material contains details on how the data were prepared and how the experiments were performed.

\subsection*{Competing financial interests}
The authors declare no competing financial interests.

%The corresponding author is responsible for submitting a \href{http://www.nature.com/srep/policies/index.html#competing}{competing financial interests statement} on behalf of all authors of the paper. This statement must be included in the submitted article file.

\end{document}